# Individual Differences in Learning Social and Non-Social Network Structures

Steve Tompson[1,2], Ari E. Kahn[1,2,3], Emily B. Falk[4,5,6], Jean M. Vettel[1,2,7], Danielle S. Bassett[1,8]

Author Note

[1]Department of Bioengineering, University of Pennsylvania

[2]Human Sciences Campaign, U.S. Army Research Laboratory

[3]Department of Neuroscience, University of Pennsylvania

[4]Annenberg School of Communication, University of Pennsylvania

[5]Department of Psychology, University of Pennsylvania

[6]Marketing Department, The Wharton School, University of Pennsylvania

[7]Department of Psychological and Brain Sciences, University of California, Santa Barbara

[8]Department of Electrical & Systems Engineering, University of Pennsylvania

Correspondence concerning this article should be addressed to Dr. Danielle S. Bassett, Department of Bioengineering, University of Pennsylvania, Philadelphia, PA 19104

Contact: dsb@seas.upenn.edu

Abstract

Learning about complex associations between pieces of information enables individuals to quickly adjust their expectations and develop mental models. Yet, the degree to which humans can learn higher-order information about complex associations is not well understood; nor is it known whether the learning process differs for social and non-social information. Here, we employ a paradigm in which the order of stimulus presentation forms temporal associations between the stimuli, collectively constituting a complex network structure. We examined individual differences in the ability to learn network topology for which stimuli were social versus non-social. Although participants were able to learn both social and non-social networks, their performance in social network learning was uncorrelated with their performance in non-social network learning. Importantly, social traits, including social orientation and perspective-taking, uniquely predicted the learning of social networks but not the learning of non-social networks. Taken together, our results suggest that the process of learning higher-order structure in social networks is independent from the process of learning higher-order structure in non-social networks. Our study design provides a promising approach to identify neurophysiological drivers of social network versus non-social network learning, extending our knowledge about the impact of individual differences on these learning processes. Implications for how people learn and adapt to new social contexts that require integration into a new social network are discussed.

*Keywords:* social network learning, statistical learning, social cognition

# Individual Differences in Learning Social and Non-Social Network Structures

Consider the important, yet daunting, challenge of learning a social network at a new job. Some connections are dictated by management structure, such as who supervises whom, project assignments, and administrative burden. Other connections may reflect personal connections from kids on the same sports team or spouses who are friends from college. Individuals may also cluster together into cliques or communities based on these individual work or personal connections. This intricate web of human interactions reflects a rich, underlying social network of relationships between individuals. Navigating these interwoven layers of social connections is critical for success at the workplace but also in social interactions with friends and family (Balkundi & Harrison, 2006; Fitzhugh & DeCostanza, 2016; Jehn & Shah, 1997; Orvis & DeCostanza, 2016). As such, understanding how people learn relational information and update social network information may provide key insights into a broad range of important questions about human behavior.

Research on statistical learning may provide insights into how people learn relational information. People are able to implicitly learn and pick up on spatial and temporal associations between objects grouped into communities (Halford, Wilson, & Phillips, 2010; Karuza, Thompson-Schill, & Bassett, 2016). Learning relational information about how objects or individuals are related to one another in space, time, or content is important for reasoning, language, and other higher cognitive processes (Halford, Wilson, & Phillips, 2010). This information enables individuals to form internal representations of the external world (Fiser & Aslin, 2002, 2005; Gómez, 2002; Saffran, Newport, & Aslin, 1996; Turk-Browne, Isola, Scholl, & Treat, 2008) which facilitate efficient information processing (Fine, Jaeger, Farmer, & Qian, 2013; Karuza, Farmer, Smith, Fine, & Jaeger, 2014; Turk-Browne, Scholl, Johnson, & Chun, 2010). By learning the relationships between objects or between individuals, people understand visual patterns, produce language (Friederici, 2005), form knowledge (Bousfield, 1953), develop social intuition (Gopnik & Wellman, 2012), exercise logical deduction, and attain expertise in their line of work (Moon, Hoffman, Novak, & Canas, 2011). Since social networks are inherently about the relations among individuals, learning relational information also likely confers advantages for successfully understanding social structure.

Collectively, relational data can be described as a network in which nodes might represent concepts, objects, or individuals, and in which edges might represent shared content, social relationships, or conditional probabilities (Moon et al., 2011). Yet, how the organization and content of such a network impacts our ability to learn the data is far from understood. Progress has been stymied by two critical limitations in both methodology and conceptualization. First, methodologically, research has predominantly focused on the learning of object pairs or concept pairs, rather than on the learning of higher-order, non-pairwise relationships present in real-world systems. Recent work suggests that human learners are sensitive to higher-order relational information beyond adjacent and immediately non-adjacent probabilities (Chan & Vitevitch, 2010; Goldstein & Vitevitch, 2014; Schapiro, Rogers, Cordova, Turk-Browne, & Botvinick, 2013). Yet, experimentally manipulating and studying these higher-order relationships requires a quantitative framework in which to characterize the network structure on which the relational data sit: that is, the arrangement of nodes and edges (Newman, 2010). The lack of such a framework has challenged our ability to predict how people might learn such higher-order relational information.

Second, conceptually, progress has been hampered by the lack of an understanding of the similarities and differences between learning relational content among objects such as visual targets or verbal commands and learning relationships among individuals, such as colleagues or friends. Previous research has studied statistical learning and social content in isolation (Wu, Gopnik, Richardson, & Kirkham, 2011). While traditional views suggest that statistical learning of relational data may be relatively agnostic to data category (symbols, syllables, visual patterns; Schapiro et al., 2013), emerging evidence demonstrates that neurobiological mechanisms are differentially recruited for learning and processing social versus non-social information (Meyer, Spunt, Berkman, Taylor, & Lieberman, 2012; Meyer, Taylor, & Lieberman, 2015). Moreover, the ability and motivation to process social information and non-social information is differentially associated with social traits, including perspective-taking (Meyer & Lieberman, 2016; Meyer et al., 2015). Thus, it remains an open question to what extent learning social and non-social relational data might rely on similar mechanisms and whether there might be unique social, cognitive, or social-cognitive factors that predict learning of social versus non-social relational data, including higher-order network structure.

Here, we addressed these methodological and conceptual challenges by studying individual differences in the learning of higher order patterns of relationships. We defined *social network learning* to be the learning of inherently social relational data embedded on a network structure. We treated objects or individuals as nodes in a network, and we treated relationships (e.g., conditional probabilities or frequencies of co-occurrence) as edges in a network. Across three studies, participants completed a basic perceptual judgment where the order in which the stimuli were presented reflected previously defined relationships between the stimuli instantiated in a clustered network architecture. The network architecture was never explicitly shown to the participants, but we hypothesized that that architecture could be inferred by the temporal associations between stimuli. More specifically, stimuli were presented such that the stimulus presented on each subsequent trial was connected in a network to the stimulus presented on the previous trial. We then manipulated the cover story for the stimuli. To study social network learning, we emphasized that the stimuli represented people; to study non-social network learning, we emphasized that the stimuli represented abstract images or rock formations (depending on the variant of the study). Importantly, we used the same visual representations across both social and non-social tasks, and only changed the meaning ascribed to the stimuli. Using this task and a post-learning categorization task, we implicitly measured the degree to which participants learned the higher order network structure of social versus non-social networks, including the cluster, or subnetwork, assignment for each image.

Using an experimental paradigm that bridges social psychology, cognitive science, and network engineering, we examined three broad questions about social and non-social network learning. First, some researchers have suggested that learning relational data operates in a manner that is independent from the type of data being learned (Schapiro et al., 2013). Thus, we hypothesized that people should learn the network structure for both social and non-social networks, and that this process should be indexed by our implicit measures of learning. Second, we asked whether there were meaningful differences in the behavioral markers of social and non-social network learning despite their broad similarities. Although people should be able to learn both social and non-social network structures, previous work has found that the processing of social information can be performed independently from the processing of non-social information (Meyer & Lieberman, 2016; Meyer et al., 2012, 2015). We therefore hypothesized that individual differences in performance on social tasks might only show weak correlations with performance on non-social tasks. Third, we investigated what traits predict social and non-social network learning. Previous work has demonstrated that processing social and non-social information is differentially associated with perspective-taking (Meyer et al., 2015), leading to our hypothesis that social traits (including perspective-taking and social orientation) should uniquely predict learning for social networks but not for non-social networks. Collectively, our results advance understanding of how people process complex relational information, and how that processing is influenced by the type of information being learned.

# Method

## Participants

We recruited a total of 349 participants across three studies. In the first two studies, we recruited participants through Amazon Mechanical Turk. Total compensation for a participant who completed all phases of either study ranged from $6.25-$9.00 (depending on performance bonuses). In Study 3, we recruited participants from the University of Pennsylvania using an online subject recruitment website (Experiments @ Penn) and compensation for the study ranged from $20-$30 (depending on performance bonuses). The protocol for all three studies was approved by the Institutional Review Board of the University of Pennsylvania.

## Experimental Design Overview

We ran a set of three behavioral experiments. We first employed a between-subjects paradigm to test for implicit signatures of network learning in social and non-social networks. In our second study, we then examined whether the group difference between social and non-social network learning could be replicated at the individual level using a within-subject design. Finally, our third study investigated whether individual differences in traits could account for variability in learning social versus non-social networks.

In all three studies, participants viewed a sequence of fractal images that we created using the Qbist filter (Loviscach & Restemeier, 2001) in the GNU Image Manipulation program (v.2.8.14; www.gimp.org), converted to grayscale, and then matched for average brightness. Each image was unique, and for each participant, each image was randomly assigned to a network node. The sequence of fractal images that each participant saw was generated by a random walk through the network (see Figure 1). This random walk ensured that the probability of one image being presented after the current trial was equivalent across trials and determined by the network structure. Each node was connected to exactly four other nodes, ensuring equivalent transition probabilities. Images were presented for 1500 ms. To ensure that participants were attending to the stream of images, they were instructed to press the J key with their right index finger if the image was rotated (30% of trials) and to press the F key with their left index finger if the image was not rotated (70% of trials). The task was broken into 5 segments and participants were given a break between segments to reduce fatigue.

To measure implicit learning of the network structure, we computed differences in RT between *pre-transition trials* that occurred immediately before a transition from one cluster to another and *post-transition trials* that occurred immediately after a transition from one cluster to another. If participants learn the higher-level network structure including cluster membership, then they should anticipate seeing a within-cluster image rather than an image from another cluster. This *surprisal effect* should slow participants' response to the rotation judgment on the next trial (Karuza, Kahn, Thompson-Schill, & Bassett, 2017; Schapiro et al., 2013). The first study also included an odd-man-out test that measured learning based on categorization of images (described below) to provide additional evidence that participants' responses were influenced by the network structure. The third study included two trait questionnaires on social orientation and perspective-taking to examine individual differences that account for variability between social and non-social network learning.

**Study 1: Between-Subjects Design to Study Social and Non-Social Network Learning**

In the first of our three studies, we used a between-subjects design to test for implicit signatures of network learning in social and non-social networks. We ran two variants of the first study on Amazon Mechanical Turk, recruiting 76 participants in the first (37 non-social, 39 social) and 82 participants in the second (40 non-social, 42 social). We excluded 5 participants (two from variant 1, three from variant 2) who had accuracy lower than chance (70%). The network structure in the first variant consisted of three clusters each composed of five nodes, and participants viewed a sequence of 1500 fractal images. In the second variant, the network structure consisted of two clusters each composed of five nodes, and participants viewed a sequence of 1000 fractal images. The purpose of this second variant was to shorten the task and to test for generalization of results across variable network size.

In the first variant, participants only read a cover story about the images in the social condition, while in the second variant, participants also read a cover story about the images in the non-social condition. The purpose of this manipulation in the second variant was to explicitly control for potential differences in cognitive load created by instructing participants to think about the images as either people or rock formations. In the social condition across both variants of Study 1, participants were told that “the images that you will see are taken from an online social media platform where people can choose one of these images as their avatar to represent themselves, much like you might use a photo to represent yourself on Facebook or Twitter. While completing the task (described in more detail on the next page), please make sure you focus on the people these avatars represent." In the second variant of Study 1, participants in the non-social condition were told that the "images were abstract patterns frequently found in rock formations. Some of these patterns are visible to the naked eye, whereas others are only visible with a microscope. These rock patterns are often created by natural forces, including tectonic plate shifts, wind and water erosion, and volcanic activity."

After performing the image rotation judgment task, participants completed an odd-man-out test. On each trial, participants were simultaneously presented with three images in random order; two of the images represented nodes next to each other in the network, and one image was drawn from nodes at least 3 connections away from one of the two original images. Participants were told that the stream of images they just saw in the exposure phase adhered to a pattern, and they were instructed to select via button-press one of the three images that "did not fit" with the other two. In each set of images, two images belonged to the same community while the third was a node outside that community. We selected groups of images such that none of the images were boundary nodes (nodes that are connected to their own community and serve as a bridge to another community), and the probability of each image being presented with other images was equivalent. Each group of images was then presented in all permuted orders giving 9 trials per group and 54 trials total in the first variant of the study (6 trials per group and 36 trials total in the second variant of the study).

Finally, in the second variant of the study, we had two additional modifications from the first: (a) a pre-exposure choice where participants were instructed to pick an image to serve as their avatar representing themselves (social condition) or to pick their favorite rock formation (non-social), and (b) a post-exposure rating task where participants reported how much they thought about the images as people on a 5-point scale. We expected that participants would report thinking about the images as people more in the social condition than in the non-social condition.

**Study 2: Within-Subjects Design to Study Social and Non-Social Network Learning**

In our second study, we complemented the between-subject approach of Study 1 with a within-subject approach. Here, we directly examined whether individuals who were better at non-social network learning were also better at social network learning. To the extent that these skills are independent, we would expect minimal relationship between performance on one task and performance on the other task. In contrast, if a common set of mechanisms underpins all types of network learning, then we would expect that performance on these two tasks would be correlated across subjects. Importantly, there could also be individual differences in motivation to learn social versus non-social networks, where some individuals are more motivated to learn social relationships than others. We ran two variants of this second study on Amazon Mechanical Turk, recruiting 65 participants in the first variant and 94 participants in the second variant. The order of the social and non-social conditions was counterbalanced across participants. We excluded five participants (one from variant 1, four from variant 2) who had accuracy lower than chance (70%).

The procedure mirrored Study 1, including the difference in cover stories between the two variants: the social cover story was present only in the first variant, while both the social and non-social cover stories were present in the second variant. The main difference from Study 1 was the number of unique fractal images used in the networks. In Study 1, we used 15 fractals that could repeat across the social and non-social network conditions. In Study 2, we used 10 unique fractals for each condition, and the images were randomly assigned to the social and non-social network for each participant. A second important difference between the two studies was that in Study 2, we did not include the odd-man out task. The third and final important difference between the two studies was that we made a modification between the two variants of Study 2. To increase the degree to which subjects differentiated between the social and non-social conditions, we instructed participants as follows in the second variant: "In this study, we are interested in how the source and context of abstract patterns influences their representation. For each part of the study, try to focus on the instructions and type of images that you are looking at IN THAT PART."

**Study 3: Influence of Social-Cognitive Traits on Social and Non-Social Network Learning**

Finally, our third study investigated whether the ability to learn the architecture of social and non-social networks is influenced by social cognitive traits. We recruited 33 participants from the University of Pennsylvania who completed the study in an on-site laboratory, and we excluded 2 participants due to missing data (server malfunction) and 1 participant who had accuracy lower than chance (70%). The procedure for Study 3 was identical to the second variant of Study 2: it included cover stories for both the social and non-social conditions, and it also included extra instructions to encourage participants to differentiate between the instructions for the two conditions. The important new feature of this study was that we asked participants to complete two questionnaires measuring individual differences in social orientation and perspective-taking.

*Social Orientation.* The Triandis Individualism-Collectivism Scale (Triandis & Gelfand, 1998) consists of 15 items measured on a 7-point scale. It is designed to assess the extent to which an individual thinks about himself or herself as independent of and distinct from others (8 items) versus the extent to which an individual thinks about himself or herself as interdependent on and connected to others (7 items). Sample independent items include, “I'd rather depend on myself than others” and “My personal identity, independent of others, is very important to me” (*M*=4.75, *SD*=0.75 α=.700). Sample interdependent items include, “I feel good when I cooperate with others” and “It is important to me that I respect the decisions made by my groups” (*M*=5.43, *SD*=0.65, α=.670). For our composite social orientation score, we reverse coded interdependent items and computed the average response across all 15 items for each participant (*M*=3.73, *SD*=0.56, α=.722).

*Perspective-Taking.* The Interpersonal Reactivity Index (Davis, 1980) consists of 28 items measured on a 5-point scale. It consists of four subscales measuring different components of empathy, including perspective-taking, fantasy, empathic concern, and personal distress. In these analyses, we focused on the most cognitive component – perspective-taking – since we did not hypothesize any involvement of fantasy or emotional responses in network learning. Sample items include, “I try to look at everybody's side of a disagreement before I make a decision” and “I sometimes try to understand my friends better by imagining how things look from their perspective”. Two of the seven items in the perspective-taking subscale were reverse coded, and we computed the average response for each participant (*M*=4.23, *SD*=0.67, α=.733).

**Data Exclusions**

To examine differences in RT due to the transition from one cluster to another, we excluded incorrect trials (8.7-11.2% data loss) and rotation trials (21.9-26.2% data loss) as well as trials with implausible response times (i.e., less than 100 ms or greater than 1500 ms; less than 1% data loss). We also excluded outlier data points greater than 3 standard deviations from the mean response time (less than 1% data loss). We also excluded a small number of trials (less than 1% data loss) where the random walk transitioned from one cluster to another and then immediately transitioned back to the first cluster, which resulted in the middle trial counting as both a pre-transition and post-transition trial. There were no significant differences in data loss across studies, or rate of data excluded for social versus non-social conditions.

**Statistical Analysis**

We tested cross-cluster differences in RT for the pre-transition and post-transition trials using the *lmer()* function (library lme4, v. 1.1-10) in R (v. 3.2.2; R Development Core Team, 2015). Slower RT on the post-transition trials that immediately follow a transition, relative to RT on the pre-transition trials that immediately precede a transition, would indicate that participants were learning the network structure (Karuza et al., 2017; Schapiro et al., 2013).

The primary mixed effects model in all three studies included node type (pre-transition versus post-transition), condition (social versus non-social), trial number (standardized), and the two-way and three-way interactions between these variables, as predictors of RT (with node type and trial number included as within-subjects variables, and with condition included as a between-subjects variable in Study 1 and within-subjects variable in Studies 2 and 3). For all models, we included the fullest set of random effects that allowed the model to converge, which included a random intercept for participant and a by-participant random slope for trial number and node type. All predictors were mean-centered to reduce multicollinearity (all $r$s<.280). We then conducted simple effects analysis to examine whether the effect of node type was significant in both the social and non-social tasks. We also ran additional analyses including repetition priming effects (number of times the image was presented in the previous 10 trials, number of trials since the image was last presented) as additional variables in a mixed effects model. Including these variables did not alter the significance of the effects reported below, and thus we focus our discussion on the first set of analyses.

Using the within-subjects design of Study 2, we were also able to test whether individuals who are better at non-social network learning were also better at social network learning. To isolate cross-cluster surprisal from individual differences in response time, we converted response times to *z*-scores (within-subject) and then computed the average difference in standardized RT for each subject. We then tested whether there was a significant correlation between the mean standardized RT difference between pre-transition and post-transition trials for social and non-social network runs. We also ran these analyses without first standardizing the response times within-subject and found the same effects when testing whether there was a significant correlation between the mean RT difference between pre-transition and post-transition trials for social and non-social network runs.

Using the additional individual differences measures collected in Study 3, we were able to test whether differences in social orientation and perspective-taking accounted for differences between network learning conditions. To examine individual differences in social and non-social network learning, we first converted RT to *z*-scores (within-subject) and computed the average standardized cross-cluster surprisal effect separately for the social task and the non-social task. We then fit linear mixed effects models with condition (social versus non-social) and scores on a single trait measure (either social orientation or perspective-taking) as predictor variables, and with standardized cross-cluster surprisal as a dependent variable.

# Results

## Confirming Attributions of Social Meaning to Fractal Images

To interpret the results of our study as relating to social versus non-social network learning, it is imperative to first demonstrate that participants attributed social meaning to the fractal images in the social condition more so than to the fractal images in the non-social condition. To address this question, we tested whether participants were significantly more likely to report thinking about the images as people in the social condition than in the non-social condition. The first study to include this measure was the second variant of Study 1, where we found that there was a significant difference in post-task ratings ($t(75.01)=3.21$, $p=.002$), such that participants reported thinking about the images as people more frequently in the social condition ($M=2.97$, $SD=1.19$) than in the non-social condition ($M=2.00$, $SD=1.48$). Consistent with the effects from Study 1, we also found a significant difference in post-task ratings in Study 2 (variant 1: $t(125.60)=2.54$, $p=.012$; variant 2: $t(173.74)=3.28$, $p=.001$), such that participants reported thinking about the images as people more frequently in the social condition (variant 1: $M=2.77$, $SD=1.16$; variant 2: $M=2.86$, $SD=1.13$) than in the non-social condition (variant 1: $M=2.22$, $SD=1.33$; variant 2: $M=2.26$, $SD=1.32$). This finding was replicated in Study 3, where we found that there was a significant difference in post-task ratings ($t(29)=2.90$, $p=.007$), such that participants reported thinking about the images as people more frequently in the social condition ($M=2.33$, $SD=0.99$) than in the non-social condition ($M=1.80$, $SD=1.06$). Collectively, these results suggest that participants were indeed more likely to think about the abstract images as people when told that they represented online avatars.

**Commonalities in Social Versus Non-Social Network Learning**

Next, we investigated whether participants were able to learn the network architecture implicit in the temporal contingencies between stimuli. Specifically, we asked whether previously identified indices of network learning in non-social domains might also index the learning of network structure in the social domain. To address this question, we examined both RT differences for pre-transition and post-transition trials as well as performance on the odd-man out task. Intuitively, slower RT on post-transition trials and greater accuracy on the odd-man out task would indicate that individuals successfully learned the network structure. We fit a linear mixed effects model with node type (pre-transition versus post-transition), condition (social versus non-social), and trial number as predictor variables, using RT as the dependent variable. Across all three studies, there was a significant main effect of node type, such that participants were significantly slower at responding to the post-transition trial than to the pre-transition trial for both social and non-social networks (see Table 1 and Figure 2). These results suggest that participants were surprised when the visual stream transitioned from one cluster to another, demonstrating that they learned the network structure of both the social and non-social networks.

A second measure of network learning is given by the participant's categorization accuracy on the odd-man out task. In this task, participants were presented with three images at a time and instructed to select one of the three images that "did not fit" with the other two. Importantly, two of the images on each trial were from the same cluster (i.e., same cluster nodes) and the third image (i.e., distant node) was at least three steps away from the other two images. Thus, if participants learned the network structure, they should be more likely to indicate that the distant node "did not fit" with the other two.

In the first variant of Study 1, participants were significantly more likely to indicate that the distant node "did not fit" with the other two same-cluster nodes in both the social task ($M$=0.415, $SD$=0.134, $t(38)$=3.95, $p$<0.001) and in the non-social task ($M$=0.386, $SD$=0.127, $t(34)$=2.61, $p$=.013), and there was no significant difference between the two conditions ($t(71.73)$=0.93, $p$=.355). We found a similar effect in the second variant, such that participants were significantly more likely to indicate that the distant node "did not fit" with the other two same-cluster nodes in the social task ($M$=0.413, $SD$=0.216, $t(37)$=2.37, $p$=.023) and marginally more likely to indicate that the distant node "did not fit" with the other two same-cluster nodes in the non-social task (*M*=0.375, *SD*=0.169, *t*(40)=1.69, *p*=.099), and there was no significant difference between the two conditions (*t*(70.05)=0.87, *p*=.386). We found the same effects in the second variant of Study 1. These results provide additional evidence that participants learned the network structure of both social and non-social networks.

**Individual Differences in Social Versus Non-Social Network Learning**

Next we turned to an examination of individual differences in social versus non-social network learning. Specifically, we were interested in determining the degree to which people who are good at learning one type of network are also good at learning the other type of network. If there was a correspondence in performance, it would suggest that the mechanism of learning social networks was similar to that of learning non-social networks. Conversely, if there was weak or no correspondence in performance, it would suggest the existence of distinct mechanisms or distinct motivations underlying social versus non-social network learning. To determine which explanation was supported by the data, we examined the correlation between each individual's cross-cluster surprisal effect in the social and non-social networks in the within-subjects data acquired from Study 2 and Study 3. We observed no correlation between learning on the social and non-social tasks for Study 2 (variant 1: *r*(62)=-0.026, p=.841; variant 2: *r*(87)=-0.157, p=.141) and Study 3 (*r*(28)=-0.074, p=.697); see Figure 3. The combined correlation across Study 2 and Study 3 was only 0.049 (*p*=.512). These data are consistent with the notion that there may be distinct processes underlying social versus non-social network learning, either in terms of motivation or in terms of learning mechanism.

To further examine the question of potentially distinct processes underlying social versus non-social network learning, we asked whether social traits of a participant predicted their ability to learn the social networks but not their ability to learn the non-social networks. We found that there was a significant interaction between social orientation and condition ($b$=-0.21, $SE$=0.08, $t$(28.00)=-2.61, $p$=.014), such that individuals who reported greater collectivistic (*versus* individualistic) cultural values showed greater cross-cluster surprisal for the social networks ($r$(28)=-0.492, $p$=.006), but there was no association between social orientation and cross-cluster surprisal for the non-social networks ($r$(28)=0.164, $p$=.387). There was also a marginally significant interaction between perspective-taking and condition ($b$=0.13, $SE$=0.07, $t$(56.00)=1.90, $p$=.063), such that individuals who reported greater perspective-taking showed greater cross-cluster surprisal for the social networks ($r$(28)=0.412, $p$=.024), but there was no association between perspective-taking and cross-cluster surprisal for the non-social networks ($r$(28)=-0.080, $p$=.674). These results suggest that people who are more in tune with others, think about the self as connected to others, and who frequently consider the perspectives of others, are more likely to learn the network structure when the network is social versus non-social. These data provide additional evidence that the learning of social networks is characterized by some processes that are independent from those implicated in the learning of non-social networks.

## Discussion

The majority of real-world systems are complex networks characterized by patterns of relationships between elements in the network (Cong & Liu, 2014; Dorogovtsev, Goltsev, & Mendes, 2008). Higher-order information about the patterns of relationships (i.e., topological features of the network) is often not captured by simply measuring pairwise associations (Barrat, Barthélemy, & Vespignani, 2008) and is an important mechanism by which people learn complex information (Chan & Vitevitch, 2010; Goldstein & Vitevitch, 2014; Halford, Wilson, & Phillips, 1998).

While there has been a recent explosion in research on topological features of complex networks across the social sciences and biological sciences (Dorogovtsev et al., 2008; Girvan & Newman, 2002; Newman, 2010), research on how people learn relational data has mostly focused on pairwise relationships without considering the type of information. Thus, it is unclear how people learn information about higher-order clustering of social information, and whether the learning process shares any similar features with previously studied processes involved in learning relational data for non-social information (Karuza et al., 2017; Qian & Aslin, 2014; Qian, Jaeger, & Aslin, 2016; Schapiro et al., 2013).

Here we show for the first time that people are capable of implicitly learning the complex, higher-order structure of social networks. Furthermore, our results suggest that social network learning may be independent from non-social network learning: we observed little correlation between individual differences in the ability to learn social versus non-social networks. Finally, social traits, including social orientation and perspective-taking, uniquely predicted learning for social networks but not for non-social networks. These results advance our understanding of how people process complex relational information, and how that processing is influenced by the type of information being learned.

**Expanding Experimental Paradigms from Non-Social to Social Network Learning**

This study extends previous work that examined statistical relationships between non-social stimuli (Fiser & Aslin, 2002, 2005; Karuza et al., 2017; Qian & Aslin, 2014; Qian et al., 2016; Schapiro et al., 2013). In this literature, statistical relationships between stimuli are represented by temporal associations (stimuli frequently presented near each other in time; Karuza et al., 2017; Qian & Aslin, 2014; Schapiro et al., 2013) or spatial associations (stimuli frequently presented at the same time; Qian et al., 2016). Individuals automatically bundle stimuli together into communities based on their temporal or spatial associations, such that stimuli that are strongly connected are processed more quickly, and people tend to respond more slowly when presented with stimuli that are not part of the current cluster (Karuza et al., 2017; Schapiro et al., 2013). Thus, individuals are capable of developing rich mental models of the higher-order topological information about the networks, even when they are not aware that such features exist (Qian et al., 2016).

Here, we observed that participants were significantly slower at responding to trials immediately following a transition from one cluster to another cluster for both social and non-social stimuli. Importantly, each node in the networks had an equivalent number of edges and thus the likelihood of moving from the pre-transition node to the post-transition node was equivalent to the likelihood of moving to any of the other within-cluster nodes that shared an edge with the current image. This architecture ensured that slower responses could not be due to differences in transition probabilities and instead is likely due to differences in cluster membership for the pre-transition and post-transition nodes. Thus, participants who responded slower to post-transition nodes had implicitly learned that the post-transition and pre-transition nodes belonged to different clusters.

Participants were also more likely to group together images that were closer together in the network, and these results did not differ for social and non-social networks, supporting the notion that participants successfully learned a higher order network structure. Again, the probability of any images being presented together in the odd-man out task was matched and all permutations were presented, and yet participants' responses suggest that they were biased by the higher-order network structure. Together, these results provide evidence for a common RT signature of network structure learning for social and non-social stimuli. Network learning for non-social stimuli plays a crucial role in cognitive performance in many other domains, including categorization, word-learning, reasoning, planning, and memory (Cong & Liu, 2014; Engelthaler & Hills, 2017; Goldstein & Vitevitch, 2014; Halford et al., 2010). It is possible that social network learning might also play a crucial role in facilitating efficient performance on social cognitive tasks such as perspective-taking, empathy, social working memory, and social reasoning.

### Dissociability of Social and Non-Social Network Learning

However, the presence of a similar RT signature of learning in social and non-social networks does not necessarily mean that the underlying processes are identical. Previous work suggests that processing social information may rely on distinct processes from processing non-social information (Gamond et al., 2012; Meyer et al., 2012; Van Overwalle, 2011; Zahn et al., 2007). For example, brain regions recruited when reasoning about other people's mental states (mentalizing) are distinct from brain regions recruited during other reasoning tasks (Van Overwalle, 2011) and brain regions involved in mentalizing predict working memory performance for social but not non-social information (Meyer et al., 2012). The ways in which people learn categories are influenced by the type of category they are learning (Ashby & Maddox, 2005, 2011; Cunningham & Zelazo, 2007) and this extends to social categories (Gamond et al., 2012). However, none of this previous work has studied complex patterns of relational information.

Our data suggest that the ability to learn social and non-social network structures are uncorrelated, and individuals who are good at learning one type of network are not necessarily good at learning the other type of network. This observation is particularly striking given that the experimental task was virtually identical except for the way in which the stimuli were described; abstract fractal-like images were randomly assigned to the social and non-social conditions, and the rotation judgment was identical across tasks. The only difference was whether the images were described as online avatars representing people or described as non-social images (abstract images in the first variants of Studies 1 and 2, and rock formations in the second variants of Studies 1 and 2 as well as Study 3). Results from a post-questionnaire confirmed the influence of the cover story where participants reported thinking about the images as people more frequently in the social condition.

Thus, it is possible that social and non-social network learning may be supported by independent processes. The strongest evidence in favor of this idea is that social, but not non-social network learning, was correlated with individual differences in perspective taking and social orientation. This observation highlights that different individuals, with different baseline motivations, performed the task differently. The lack of an interaction between node type (pre- vs. post-transition) and condition (social vs. non-social) in all but one of our study variants, however, leaves open the possibility that the underlying mechanisms may overlap and be called upon according to these differing motivational forces. Additional research is needed to disentangle these different possible interpretations.

We also found some evidence that the rate at which participants learn the social versus non-social stimuli also differs. In Study 1, participants demonstrated smaller cross-cluster surprisal effects at the beginning of the social network learning task (versus non-social network learning task) but this difference between social and non-social diminished over time, such that the cross-cluster surprisal effects were equivalent at the end of the task (see Table 1).

There are two plausible interpretations of this effect. First, this effect could be due to social network learning repurposing network learning of non-social information, much like other types of social cognition modify and repurpose other "ancestral" cognitive processes (Immordino-Yang, Chiao, & Fiske, 2010; Parkinson & Wheatley, 2015). This process might involve scaffolding of the social information on top of basic processing and would result in increased task demands and slower learning of the social network structure. Second, social information could actually be processed first and could instead bias the subsequent processing of detail. Both scenarios would lead to slower learning of the social network structure. However, it is important to note that the interaction between node type and time was only present in the between-subjects designs in Study 1, and the interaction was not significant in the within-subjects paradigms used in Studies 2 and 3. It is possible that the order in which the participants saw the two networks, or the fact that they saw both networks, obfuscated the interaction between node type and time, although further work is needed to directly test this possibility.

**Social Traits Uniquely Predict Social Network Learning**

Another important test of the similarities (or differences) in learning social vs. non-social network structure concerns the trait-level predictors of social and non-social network learning. To the extent that learning social networks and learning non-social networks involve independent processes, we would expect them to be predicted by different traits. Consistent with this hypothesis, we find that perspective-taking and social orientation uniquely predict social network learning but not non-social network learning. Thus, individuals who are more likely to consider the mental states of others and think about the self as being closely connected to others are more likely to learn the higher-order structure of the social networks.

People who are high in collectivistic social orientation are more likely to be concerned with social relationships and maintaining social harmony (Kim & Markus, 1999; Markus et al., 1991; Tompson, Lieberman, & Falk, 2015; Triandis & Gelfand, 1998), and may therefore be more likely to pick up on relational information in social networks. Moreover, people from collectivistic cultures are more likely to attend to contextual information (Chua, Boland, & Nisbett, 2005; Nisbett, Peng, Choi, & Norenzayan, 2001) and perceive relationships in the environment (Ji, Peng, & Nisbett, 2000).

This work also extends previous evidence suggesting that individual differences in ability to maintain social information in working memory is uniquely predicted by perspective-taking, whereas no such relationship exists for working memory for non-social information (Meyer & Lieberman, 2016; Meyer et al., 2012, 2015). We build on this earlier work to show that learning of social networks is also uniquely predicted by perspective-taking, and expand it to show that other social traits including social orientation also predict social network learning.

**Real World Applications**

Understanding how people learn complex social networks has important implications for many real-world domains. In fact, the majority of real-world systems can be described by complex patterns of relationships between elements in the network (Cong & Liu, 2014; Dorogovtsev, Goltsev, & Mendes, 2008). Furthermore, network structure is a key driver of group behavior and has been studied in the context of environmental disasters (Bosworth & Kreps, 1986), terrorist networks (Krebs, 2002), gangs (Van Gennip et al., 2013), and many other social and biological systems (Girvan & Newman, 2002). In an increasingly mobile world, people are frequently interacting, living, and working with novel groups of people. To successfully adapt to these new contexts and integrate into new communities, it will be crucial for individuals to learn the social dynamics of that community, including learning the higher-order topological information about that network.

**Methodological Considerations**

One potential limitation of the current work is that data for Studies 1 and 2 were collected online through MTurk. This collection method allows for rapid collection of large samples of survey and behavioral data, but also introduces noise into the study. Although MTurk participants are at least as attentive as participants drawn from college samples, there are risks associated with collecting data from a pool of participants who might complete dozens of surveys and experiments per month (Chandler, Mueller, & Paolacci, 2014; Crump, McDonnell, Gureckis, Romero, & Morris, 2013; Hauser & Schwarz, 2016).

Moreover, our primary measure across all of the experiments was RT, which is likely influenced by variability in the computer, web browser, and internet quality used by each participant. However, our primary dependent variable focused on within-participant variability in RT, and thus any concerns about between-subject variability in RT are mitigated. Moreover, in Study 3 we recruited participants from the community around Philadelphia and had them complete the experiment in a laboratory under controlled experimental conditions. The mean RT (range=749 to 751 ms), accuracy (range=89.8% to 91.3%), and cross-cluster surprisal effect (range= 24 to 29ms) were very similar in the MTurk samples and community sample. Converging evidence across Studies 1 and 2 (MTurk samples) and Study 3 (community sample) helps to strengthen our confidence in these findings.

Another potential limitation is the small set of stimuli and single network topology. In order to demonstrate a clear effect with minimal variation across social and non-social networks, we chose to focus our experiments on abstract images chosen from a small set and only examined two network configurations with very clear clusters. It is therefore unclear whether the effects described here might be influenced by the topology, such that it might be more difficult to learn more complex network topologies or network topologies with more transition edges between communities.

Additionally, we did not test non-social traits. Given that social traits uniquely predicted social network learning, one potential hypothesis is that non-social network learning should be uniquely predicted by non-social traits (including working memory ability, intelligence, etc.). However, if social information processing scaffolds on top of basic cognitive processing, then it is also possible that non-social cognitive abilities might influence both social and non-social network learning, even though social traits only influence social network learning. Future work should include additional measures of non-social traits to test these competing hypotheses.

## Conclusion

In this paper, we discussed statistical learning of social and non-social network structures. Statistical learning is an important process whereby people learn the relationship between features or pieces of information based on their frequency of occurring near each other in space or time (Fiser & Aslin, 2002, 2005). While this topic has been heavily studied in the non-social domain (Karuza et al., 2017; Qian & Aslin, 2014; Qian et al., 2016; Schapiro et al., 2013), no research to date has examined this process in the social domain. However, it is likely that statistical learning plays a crucial role in learning social networks, such as when individuals start a new job or encounter a new social group. Taken together, these results suggest that individuals are able to learn the higher-order network structure of both social and non-social information. Importantly, although there are similarities in the implicit learning signatures, there also appear to be distinct processes involved in learning social and non-social network structures. These results advance understanding of how people build mental models of both social and non-social features of the natural world. This research has important implications for how quickly people will learn and adapt to new social contexts that require integration into a new social network. Future research should examine whether individual differences in these abilities are linked to psychological adjustment and well-being following a move or social transition.

Acknowledgements

This work was supported by an NSF CAREER award to DSB (CAREER PHY-1554488) and by an award from the Army Research Laboratory (W911NF-10-2-0022) to support collaboration between DSB, EBF, and JMV. DSB would also like to acknowledge support from the John D. and Catherine T. MacArthur Foundation, the Alfred P. Sloan Foundation, the Army Research Office through contract number W911NF-14-1-0679, the National Institute of Health (2-R01-DC-009209-11, 1R01HD086888-01, R01-MH107235, R01-MH107703, R01MH109520, 1R01NS099348 and R21-M MH-106799), the Office of Naval Research, and the National Science Foundation (BCS-1441502, BCS-1631550, and CNS-1626008). EBF would also like to acknowledge support from NIH 1DP2DA03515601, DARPA YFA D14AP00048 and HopeLab. JMV acknowledges support from mission funding to the U.S. Army Research Laboratory. We would also like to thank Elisabeth A. Karuza for helpful feedback on the study design and manuscript. The content is solely the responsibility of the authors and does not necessarily represent the official views of any of the funding agencies.

Table 1.

*Summary of results after fitting a mixed effects model.*

| | Study 1 (Variant 1) | Study 1 (Variant 2) | Study 2 (Variant 1) | Study 2 (Variant 2) | Study 3 |
|---|---|---|---|---|---|
| *All Trials* | | | | | |
| Main Effect of Node Type | **b=15.05, SE=1.51, t(68)=9.99, p<0.001** | **b=11.91, SE=2.26, t(4,117)=5.28, p<0.001** | **b=14.52, SE=1.29, t(67)=11.29, p<0.001** | **b=14.33, SE=1.16, t(87)=12.39, p<0.001** | **b=8.53, SE=2.51, t(357)=3.40, p<0.001** |
| Main Effect of Condition | b=11.95, SE=8.63, t(72)=1.38, p=.171 | b=15.61, SE=18.85, t(77)=0.83, p=.410 | b=.01, SE=1.25, t(12,170)=0.01, p=.992 | b=-0.21, SE=1.04, t(17,757)=-0.20, p=.839 | b=-5.93, SE=3.47, t(5,946)=-1.71, p=.087 |
| Main Effect of Trial Number | **b=-37.34, SE=2.82, t(72)=-13.23, p<0.001** | **b=-39.36, SE=4.41, t(77)=-8.92, p<0.001** | **b=-24.44, SE=2.32, t(63)=-10.54, p<0.001** | **b=-25.61, SE=2.09, t(88)=-12.23, p<0.001** | **b=-28.01, SE=3.39, t(54)=-8.26, p<0.001** |
| Node x Condition Interaction | b=-0.77, SE=1.51, t(68)=-0.51, p=.609 | b=2.45, SE=3.29, t(4,200)=0.74, p=.457 | b=1.88, SE=1.24, t(12,130)=1.51, p=.131 | **b=-2.87, SE=1.04, t(17,741)=-2.77, p=.006** | b=3.54, SE=3.45, t(5,924)=1.03, p=.305 |
| Node x Trial Interaction | b=1.81, SE=1.39, t(10,271)=1.30, p=.194 | b=-1.86, SE=2.25, t(7,414)=-0.83, p=.409 | b=1.37, SE=1.24, t(12,160)=1.10, p=.270 | **b=2.20, SE=1.04, t(17,775)=2.12, p=.034** | b=0.04, SE=2.44, t(5,930)=0.02, p=.985 |
| Condition x Trial Interaction | b=-5.01, SE=2.82, t(72)=-1.77, p=.080 | **b=14.04, SE=6.38, t(78)=2.20, p=.031** | **b=2.48, SE=1.25, t(12,170)=1.99, p=.047** | b=-1.35, SE=1.04, t(17,797)=-1.30, p=.195 | **b=7.54, SE=3.48, t(5,937)=2.17, p=.030** |
| Node x Condition x Trial Interaction | **b=4.68, SE=1.39, t(10,271)=3.36, p<0.001** | **b=7.36, SE=3.29, t(7,415)=2.24, p=.025** | b=-.05, SE=1.24, t(12,160)=-0.41, p=.685 | b=-1.11, SE=1.04, t(17,762)=-1.08, p=.282 | b=0.14, SE=3.45, t(5,927)=0.04, p=.967 |
| *Non-Social Network* | | | | | |
| Main Effect of Node Type | **b=15.91, SE=2.23, t(34.04)=7.12, p<0.001** | **b=11.83, SE=2.27, t(546)=5.20, p<0.001** | **b=12.625, SE=1.77, t(64)=7.14, p<0.001** | **b=17.13, SE=1.51, t(432)=11.33, p<0.001** | **b=8.75, SE=2.63, t(52.94)=3.32, p=.002** |
| Main Effect of Trial Number | **b=-32.29, SE=4.52, t(34.15)=-7.14, p<0.001** | **b=-39.42, SE=4.22, t(41)=-9.33, p<0.001** | **b=-27.53, SE=3.35, t(61)=-8.22, p<0.001** | **b=-24.84, SE=2.71, t(89)=-9.16, p<0.001** | **b=-28.39, SE=3.56, t(30.03)=-7.98, p<0.001** |
| Node x Trial Interaction | b=-2.94, SE=2.05, t(207.34)=-1.44, p=.152 | b=-1.86, SE=2.24, t(3,950)=-0.83, p=.407 | b=1.38, SE=1.71, t(5,989)=0.81, p=.418 | **b=2.92, SE=1.44, t(8,831)=2.03, p=.042** | b=-0.12, SE=2.86, t(23.82)=-0.04, p=.967 |
| *Social Network* | | | | | |
| Main Effect of Node Type | **b=14.25, SE=2.10, t(40.57)=6.78, p<0.001** | **b=14.32, SE=2.50, t(245)=5.74, p<0.001** | **b=16.14, SE=1.74, t(61)=9.30, p<0.001** | **b=11.78, SE=1.50, t(85)=7.87, p<0.001** | **b=12.14, SE=2.72, t(43.99)=4.47, p<0.001** |
| Main Effect of Trial Number | **b=-42.41, SE=3.47, t(37.14)=-12.21, p<0.001** | **b=-25.29, SE=4.83, t(37)=-5.24, p<0.001** | **b=-23.29, SE=2.88, t(59)=-8.09, p<0.001** | **b=-27.90, SE=2.76, t(85)=-10.11, p<0.001** | **b=-22.78, SE=4.55, t(29.46)=-5.01, p<0.001** |
| Node x Trial Interaction | **b=6.56, SE=2.01, t(211.13)=3.27, p=.001** | **b=5.54, SE=2.41, t(3,483)=2.30, p=.021** | b=1.43, SE=1.70, t(6,050)=0.84, p=.399 | b=1.04, SE=1.43, t(8,770)=0.73, p=.469 | b=0.13, SE=2.78, t(35.18)=0.05, p=.964 |

*Note.* Significant effects are shown in bold.

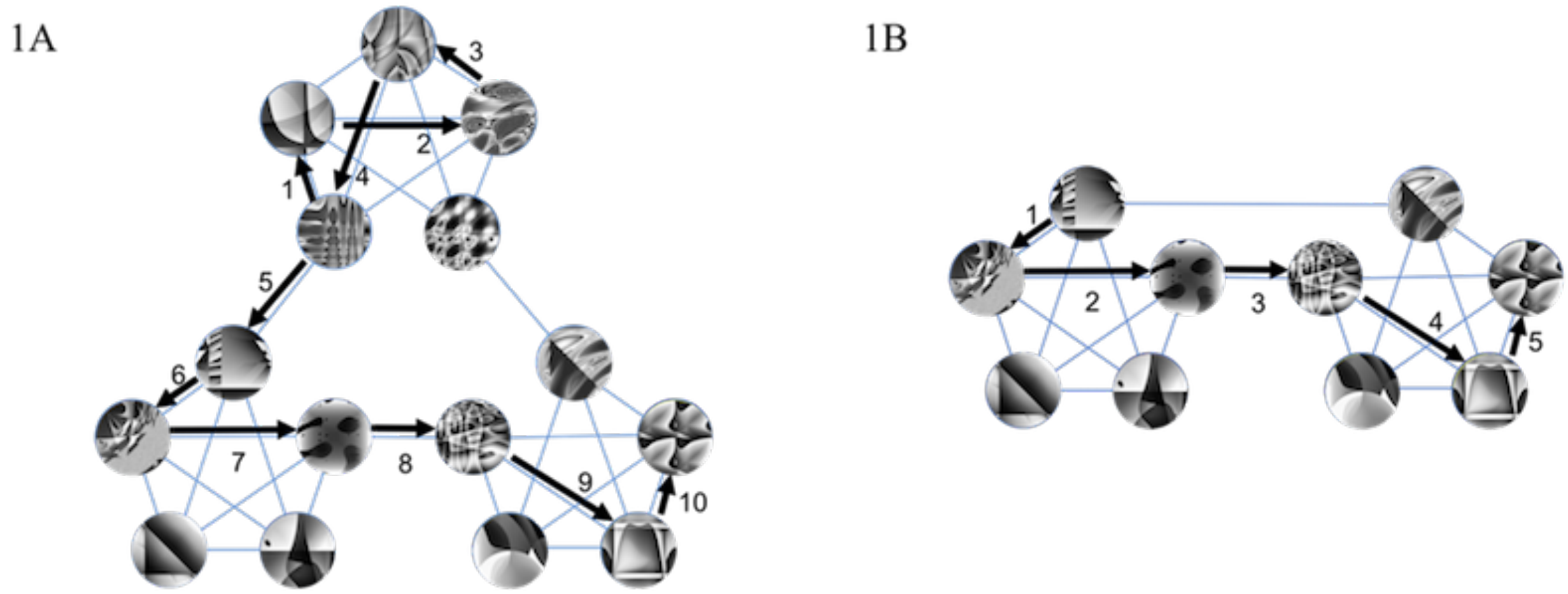

*Figure 1*. Random walk through network of fractal images. The first version of Study 1 consisted of a random walk through three clusters of five images (Figure 1A) whereas all other studies consisted of a random walk through two clusters of five images (Figure 1B).

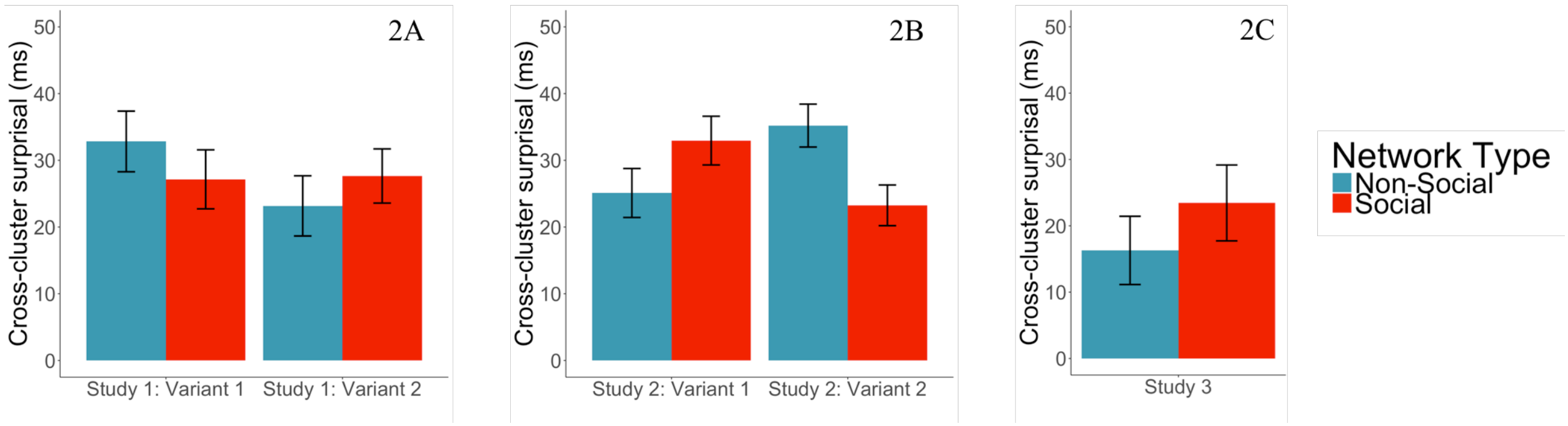

*Figure 2.* Difference in RT for post-transition minus pre-transition trials for social and non-social networks. In all three studies and both social and non-social networks, participants responded significantly slower on post-transition trials than on pre-transition trials.

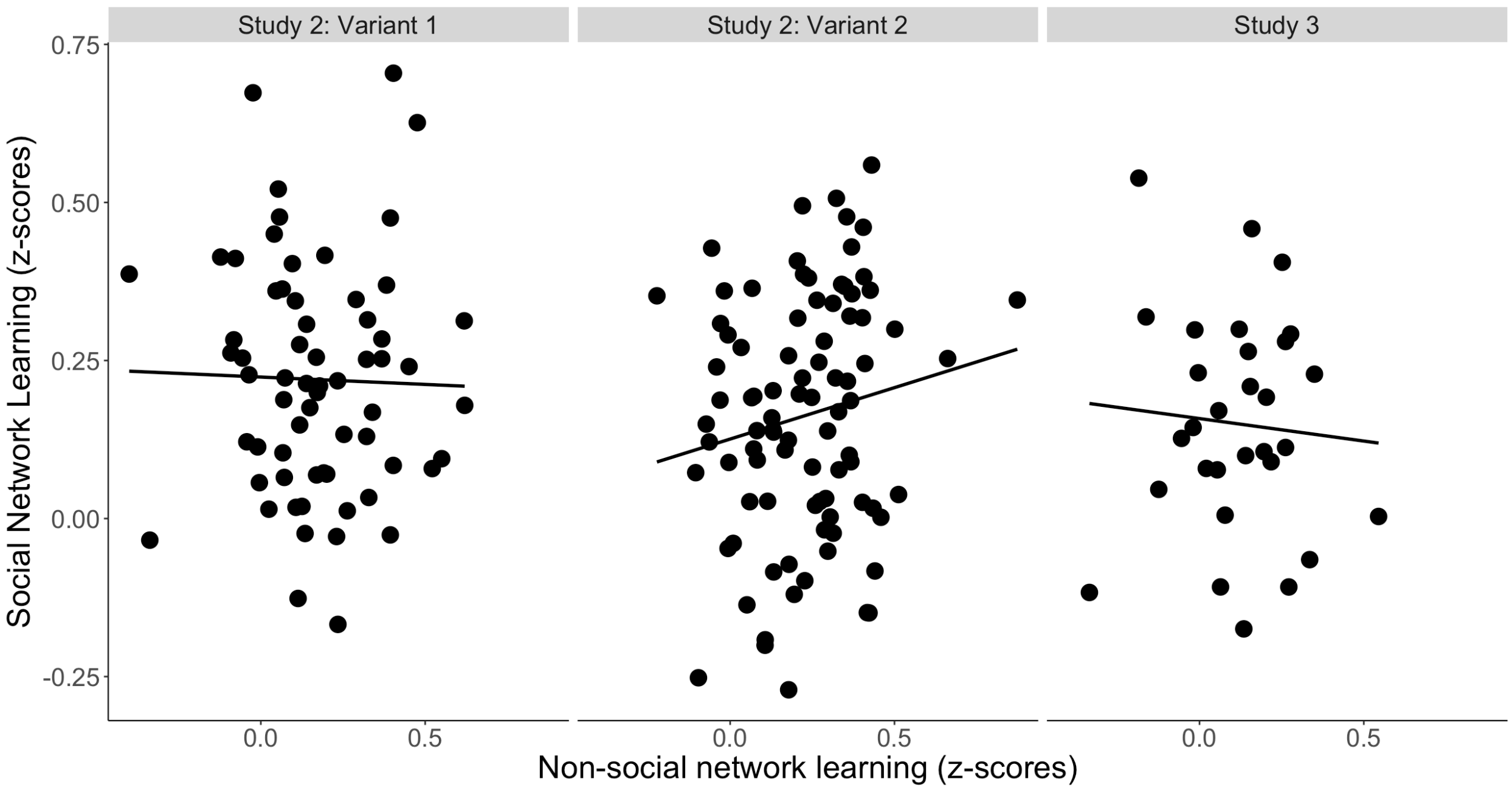

*Figure 3.* Correlation between each individual's cross-cluster surprisal effect (standardized within subject) for the social network and non-social network conditions. There were no significant associations between social and non-social network learning in Study 2 or in Study 3.

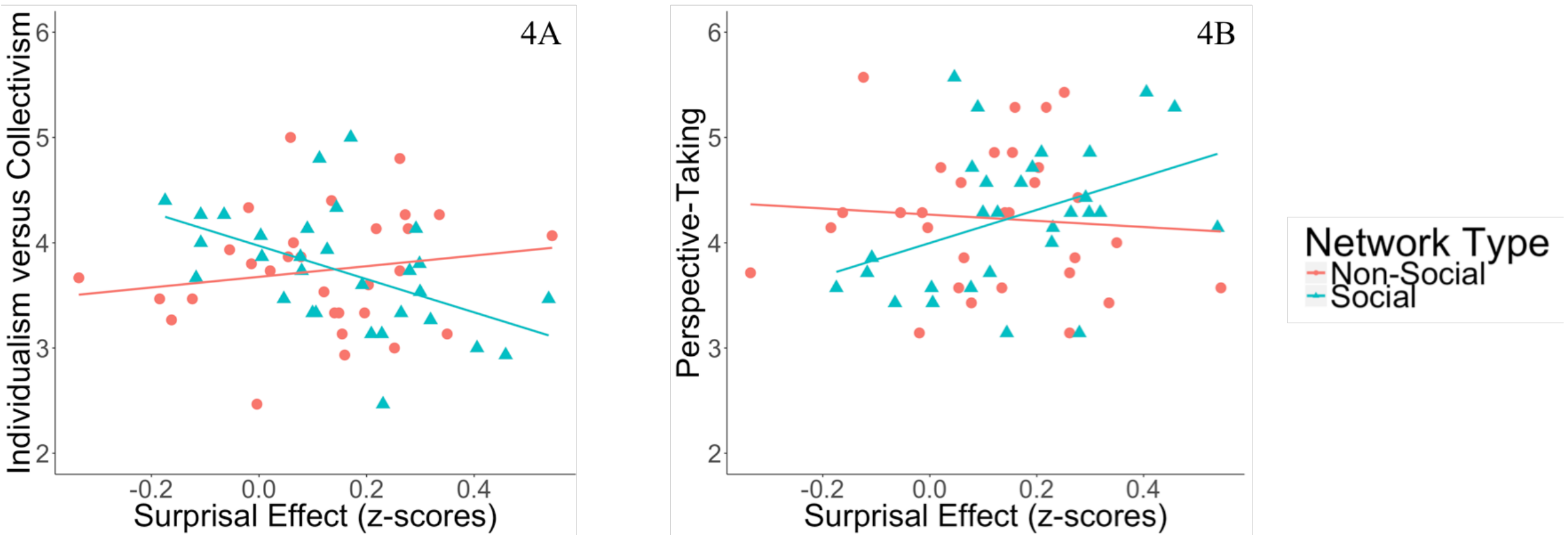

*Figure 4.* Association between social traits and cross-cluster surprisal. People who are more collectivistic (Fig 4A) and people who are more likely to consider the perspective of others (Fig 4B) show stronger cross-cluster surprisal for the social networks but not for the non-social networks.